\documentclass[12pt]{iopart}
\usepackage{iopams}  

\usepackage{cite}
\usepackage[colorlinks, citecolor = blue, urlcolor = blue]{hyperref}
\expandafter\let\csname equation*\endcsname\relax
\expandafter\let\csname endequation*\endcsname\relax

\usepackage{amssymb}
\usepackage{amsmath}
\usepackage{soul}

\usepackage{enumerate}
\usepackage{empheq}
\usepackage{fancybox}
\usepackage{etoolbox}
\usepackage{hyperref}
\usepackage{float}
\usepackage{color}
\usepackage{orcidlink}
\makeatletter
\def\@mkboth#1#2{}
\newlength\appendixwidth
\preto\appendix{\addtocontents{toc}{\protect\patchl@section}}
\newcommand{\patchl@section}{%
	\settowidth{\appendixwidth}{\textbf{Appendix }}%
	\addtolength{\appendixwidth}{1.5em}%
	\patchcmd{\l@section}{1.5em}{\appendixwidth}{}{\ddt}%
}
\makeatother
\begin{document}
	
	\title{Convergence to the  Solution of the One-Dimensional Riviera Model} 

	\author{Pascal Viot\orcidlink{0000-0002-1578-1722}}
	\address{Sorbonne Université, Laboratoire de Physique Théorique de la Matière Condensée, UMR CNRS 7600, 4, Place Jussieu, 75005 Paris, France}
\date{\today}
\begin{abstract}
The Riviera model is a sequential random placement process in which a new house cannot be built if doing so would result in a configuration with three consecutive houses. In other words, a new house can only be built if its location is surrounded by empty adjacent sites or by a single occupied neighbor that is itself isolated. Despite the apparent simplicity of this rule, no exact analytical solution of the original model is currently known. We derive an infinite hierarchy of kinetic equations governing its dynamics and introduce a systematic closure scheme that can be implemented at arbitrary finite order. Solving the resulting systems order by order, we show that the corresponding kinetics converges rapidly toward the solution. At high order, the method yields an exceptionally accurate estimate of the jamming density while retaining explicit analytical expressions for the full time-dependent evolution.
\end{abstract}

\submitto{J. Stat. Mech.}

\section{Introduction}

Irreversible filling processes provide a natural setting in which simple local rules generate nontrivial collective behavior. They arise in a broad range of problems in mathematics, statistical physics, and materials science, from classical packing questions to random adsorption and deposition processes \cite{Torquato10,Chen2014,Jiao2009,Hales05,doslic2020}. Whereas packing theory is primarily concerned with the highest density attainable under prescribed geometrical constraints \cite{Conway}, random sequential processes address a complementary question: how does a configuration evolve when objects are deposited one after another according to local acceptance rules? Such processes appear in contexts ranging from molecular biology, immunology, and genomics \cite{Ewens2005,Athreya04} to topological data analysis \cite{Bobrowski17,Tillmann22,Leykam2023}, ballistics \cite{Hall88}, number theory \cite{Rogers,GL87,Kannan1987}, and stochastic optimization \cite{ZZ2008,Gomez2019}.

Settlement-planning models belong to this broad class of constrained filling processes. In these models, houses are built sequentially on a lattice under geometrical or environmental restrictions, such as prescribed access to sunlight or open space \cite{Puljiz2022}. The dynamics terminates in a saturated configuration in which no further house can be added, and the principal observable is the resulting fraction of occupied land. Exact enumeration and numerical approaches have been successfully applied to finite two-dimensional systems of small or moderate size \cite{Puljiz2022,Puljiz2023}. Related variants have also been introduced recently \cite{Puljiz2025}, and similar exclusion mechanisms occur in other settings, including seating models with social-distancing constraints \cite{Spahn2025}.

The one-dimensional version of this settlement problem, known as the Riviera model, was introduced in \cite{Doslic2024} and subsequently investigated in \cite{Luck2024}. Deposition attempts are made uniformly at random on a one-dimensional lattice, and a house is built only if doing so does not result in a configuration of three consecutive houses. Although this rule is close in spirit to the Flory model for the random sequential adsorption of dimers \cite{Flory39}, the Riviera model is considerably more difficult to analyze. The essential difference is the absence of the screening property that renders disconnected empty intervals statistically independent in many exactly solvable one-dimensional adsorption models \cite{Renyi58,Evans93,Talbot2000}. Consequently, the dynamics generates correlations over increasingly long distances, and the exact evolution cannot be reduced to a finite closed set of kinetic equations.

A closely related modification was recently shown by Krapivsky \cite{Krapivsky2026} to possess a finite correlation range and to be exactly solvable. This comparison makes clear that the analytical difficulty of the original Riviera model is directly tied to the  dynamical correlations produced by its deposition rule. The original model therefore provides a particularly instructive example of a stochastic process that is elementary to formulate but lies beyond the reach of standard exact methods.

In the present work, we show that the kinetics of the original one-dimensional Riviera model can nevertheless be described with very high accuracy through a systematic hierarchy of approximations. Starting from the exact evolution equations for joint occupation probabilities, we close the hierarchy at an arbitrary finite order by replacing  largest intervals with mean-field expressions. Each truncation yields a finite linear system that can be solved exactly, thereby providing explicit analytical expressions for the complete temporal evolution and increasingly accurate estimates of the jamming density.

A central result of this study is the remarkably rapid convergence of the hierarchy with the truncation order. Low-order closures capture only the qualitative structure of the dynamics, whereas higher-order approximations progressively incorporate the correlations generated over larger spatial scales and quickly approach the asymptotic behavior. Beyond the Riviera model itself, the combination of exact hierarchy equations, systematic closure, and symbolic computation provides a general strategy for studying irreversible deposition processes that do not admit an exact finite-dimensional description.

\section{The model}\label{sec:model}
We consider a finite one-dimensional lattice of $L$ sites with periodic boundary conditions. Each site $i$ is characterized by an occupation variable $n_i\in\{0,1\}$, where $n_i=1$ indicates that the site is occupied by a house, while $n_i=0$ corresponds to a vacant site. The system is initially empty, $n_i(0)=0$ for all $i$.

\subsection{Definition of the model}

We consider a finite one-dimensional lattice of $L$ sites with
periodic boundary conditions. Each site $i$ is characterized by an
occupation variable $n_i\in\{0,1\}$, where $n_i=1$ indicates that
the site is occupied by a house, while $n_i=0$ corresponds to a
vacant site. The lattice is initially empty, i.e., $n_i(0)=0$
for all $i$.\\
The system evolves according to a continuous-time Markov process
describing irreversible deposition. Each site is independently
selected for a deposition attempt at unit rate. A house occupying
a single site can be deposited at a vacant site $i$ provided that
the newly deposited house has at least one vacant nearest neighbor
and that every previously occupied neighboring site retains at
least one vacant nearest neighbor after deposition.
Otherwise, the deposition attempt is rejected.

Denoting the configuration of the system by
$\boldsymbol{n}=(n_1,\ldots,n_L)$, the transition rate for
depositing a house at site $i$ is given by
\begin{equation}
	\label{eq:transition_rate}
	w_i(\boldsymbol{n}) =
	(1-n_i)
	(1-n_{i-2}n_{i-1})
	(1-n_{i-1}n_{i+1})
	(1-n_{i+1}n_{i+2}),
\end{equation}
where site indices are understood modulo $L$.

The corresponding Markov generator $\mathcal{L}$, acting on an
arbitrary observable $f(\boldsymbol{n})$, reads
\begin{equation}
	\label{eq:markov_generator}
	(\mathcal{L}f)(\boldsymbol{n})
	=
	\sum_{i=1}^{L}
	w_i(\boldsymbol{n})
	\left[
	f(\boldsymbol{n}^{\,i})-f(\boldsymbol{n})
	\right],
\end{equation}
where $\boldsymbol{n}^{\,i}$ denotes the configuration obtained
from $\boldsymbol{n}$ by setting $n_i=1$, leaving all other
occupation variables unchanged.

The dynamics proceeds until the system reaches an absorbing
configuration in which no further deposition is possible.
Such a configuration, referred to as a jammed state, satisfies
\begin{equation}
	\label{eq:jammed}
	w_i(\boldsymbol{n})=0,
	\qquad i=1,\ldots,L.
\end{equation}
\begin{figure}[H]
	\begin{center}
		\includegraphics[scale=0.45]{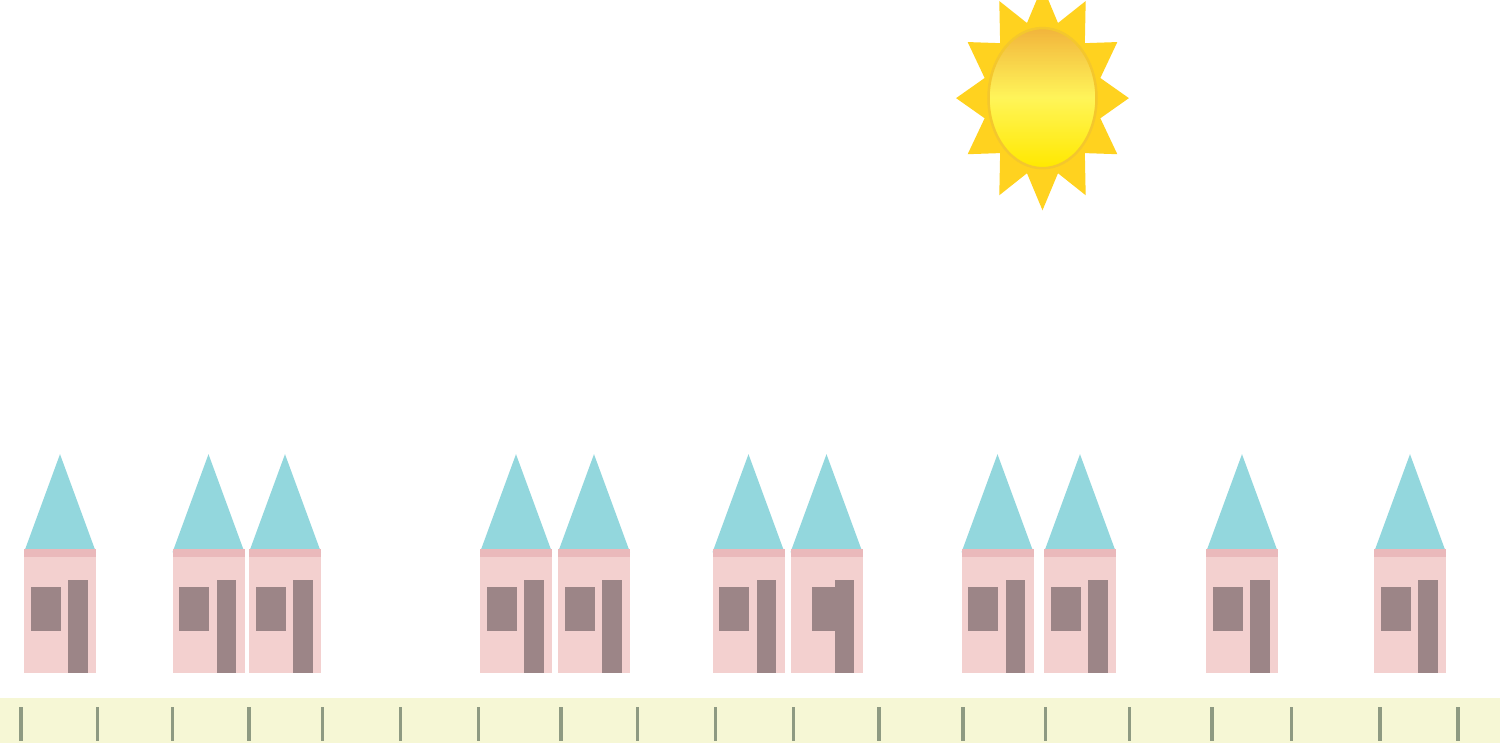}		
	\end{center}
	\caption{Example of a saturated configuration of the one-dimensional Riviera model.} \label{fig:riviera}
\end{figure}

In particular, the alternating configuration
$\ldots 101010\ldots$ and the periodic dimer configuration
$\ldots 11001100\ldots$ are both jammed states. Figure~\ref{fig:riviera} shows a typical saturated configuration of the Riviera model consisting in monomers, dimers, single or double vacant sites.
More generally, the jammed configurations dynamically accessible
from the initially empty lattice consist of isolated houses and
pairs of adjacent houses, separated by one or two vacant sites,
with no further deposition allowed.

Each site is either empty or occupied. We denote by $\pi_j(t)$, with $j=0,1$, the probability that a randomly chosen site is in state $j$ at time $t$. Thus, $\pi_0(t)$ and $\pi_1(t)$ are the densities of empty and occupied sites, respectively. Probability conservation gives
\begin{equation}
	\sum_{j=0}^1 \pi_j(t)=1
\end{equation}

Their time evolution is determined by gain and loss terms involving the joint probabilities
$p_{i_1 i_2\ldots i_l}(t)$ of observing a prescribed sequence of $l$ consecutive sites where $i_j\in {0,1}$. For instance, $p_{000}$ is the probability of finding a string
	of $3$ empty sites or empty $3$-tuple (which could be part of
	a longer string).
	
At arbitrary order, the left--right symmetry of the model yields 
\begin{equation}\label{eq:sumrules}
	p_{i_1,i_2,...,i_l}=p_{i_l,i_{l-1},...,i_1}
\end{equation}

The joint probabilities also satisfy the consistency relations
\begin{equation}\label{eq:rules}
	\sum_{i_l}p_{i_1,i_2,...,i_l}=p_{i_1,i_2,...,i_{l-1}}
\end{equation} 

The time evolution of 
 the one-site densities obeys \cite{Cohen1963,Evans93,Bonnier1994,Talbot2000,Redner2010}
\begin{equation}\label{eq:riviera}
	\begin{split}
		\frac{d\pi_0(t)} {dt}\equiv \frac{dp_0}{dt}&= -p_{000}-2p_{0010}\\
		\frac{d\pi_1(t)}{dt}\equiv \frac{dp_1}{dt}&=p_{000}+2p_{0010}  
	\end{split}
\end{equation}

The empty-site density decreases, and the occupied-site density increases, whenever a deposition occurs either at the central site of a sequence $000$ or at the appropriate site of a sequence $0010$ or its mirror image. The factor of $2$ in Eq.~\eqref{eq:riviera} accounts for these two mirror-related configurations. 

The same joint probabilities determine the densities of monomers and dimers. Here, a monomer is an occupied site with two empty nearest neighbors, whereas a dimer is a pair of adjacent occupied sites bounded by empty sites. Their densities, denoted by $\pi_m(t)$ and $\pi_d(t)$, obey
\begin{equation}\label{eq:md}
	\begin{split}
		\frac{d\pi_m(t)} {dt}\equiv \frac{dp_{010}}{dt}&= p_{000}-2p_{0010}\\
		\frac{d\pi_d(t)}{dt}\equiv \frac{dp_{11}}{dt}&=2p_{0010}  	
	\end{split}
\end{equation}

As expected, the total occupied-site density is the sum of the monomer and dimer contributions: 
\begin{equation}
	\pi_1(t)=\pi_m(t)+2\pi_d(t)
\end{equation}

To determine $\pi_0(t)$ and $\pi_1(t)$, one must therefore obtain the time evolution of the two key joint probabilities $p_{000}(t)$ and $p_{0010}(t)$. Their exact evolution equations are

\begin{equation}\label{eq:empty3}
	\begin{split}
		\frac{dp_{000}(t)} {dt}&= -p_{000}-2p_{00010}-2p_{0000}\\
		\frac{dp_{0010}(t)}{dt}&= -p_{0010}-p_{00010}-p_{010010}-p_{00100}+p_{0000}
	\end{split}
\end{equation}
For $p_{000}(t)$, the loss terms describe deposition at the central site and deposition events involving either boundary of the three-site interval. For $p_{0010}(t)$, the four loss terms account for all admissible depositions that destroy the sequence, while the gain term arises when a deposition in a four-site empty interval creates the pattern $0010$. 

Because the deposition rule probes sites up to distance two from the target, the evolution of a probability defined on an interval of length $n$ generally depends on probabilities associated with intervals of lengths $n+1$ and $n+2$. Equation~\eqref{eq:empty3} illustrates this hierarchical structure. At the first level, the one-site densities require $p_{000}$ and $p_{0010}$. At the next level, the evolution of these quantities involves $p_{00010}$, $p_{0000}$, $p_{010010}$, and $p_{00100}$. The exact kinetic description therefore contains an infinite hierarchy of probabilities associated with intervals of increasing length.

In a jammed configuration, every remaining empty site is isolated and has at least one occupied nearest neighbor on each side of the corresponding forbidden deposition environment. Simple periodic arrangements provide bounds on the occupied-site density. An alternating pattern of occupied and empty sites gives $\pi_1=1/2$, whereas a periodic arrangement of dimers separated by a single vacancy gives $\pi_1=2/3$. The accurate numerical estimate reported in \cite{Krapivsky2023a} is $\pi_1\simeq 0.600385$. The jammed state generated by the stochastic dynamics is therefore a nontrivial mixture of monomers and dimers. 	

\section{Closure of the hierarchy}
The occupation densities are determined by joint probabilities of increasingly long sequences because every admissible deposition event depends on the local environment of the target site. As noted above, the evolution equation for a probability defined on $k$ sites involves probabilities defined on $k+1$ and $k+2$ sites. The exact kinetic equations therefore do not close at any finite order.

We construct a systematic approximation by truncating the hierarchy at a prescribed interval length $l$. Probabilities associated with intervals of length $l$ or greater are replaced by mean-field expressions that retain the correct initial condition and a simple exponential time dependence. Increasing $l$ progressively incorporates correlations over larger spatial scales and is expected to improve the approximation to the exact kinetics.

The closure used throughout this work has the generic form
\begin{equation}
	p_{i_1,i_2,...,i_l}=e^{-mt}(1-e^{-pt})
\end{equation}
where $m$ denotes the number of empty sites in the interval and $p$ the number of occupied sites. Although this prescription neglects correlations beyond the truncation scale, it preserves the algebraic structure of the hierarchy (namely, the solutions obtained by solving the ODE are always a finite combination of exponentials) and leads to finite systems of linear differential equations that can be solved exactly.

\subsection{First-order}
At first order, the hierarchy is closed directly at the level of the two probabilities entering Eq.~\eqref{eq:riviera}. We use the ansatz  
\begin{equation}
	\begin{split}
	p_{000}^{(1)}&=e^{-3t}\\
	p_{0010}^{(1)}&=e^{-3t}(1-e^{-t})
	\end{split}
\end{equation}
where the superscript $(1)$ denotes the first-order approximation. 

Solving then Eqs.~\eqref{eq:riviera} and \eqref{eq:md}, one obtains
	\begin{equation}
			\begin{split}
	\pi_0(t)&=-\frac{{\mathrm e}^{-4 t}}{2}+{\mathrm e}^{-3 t}+\frac{1}{2}\\
	\pi_m(t)&=
	-\frac{{\mathrm e}^{-4 t}}{2}+\frac{{\mathrm e}^{-3 t}}{3}+\frac{1}{6}\\
	\pi_d(t)&= \frac{{\mathrm e}^{-4 t}}{2}-\frac{2 {\mathrm e}^{-3 t}}{3}+\frac{1}{6}\\
	\pi_1(t)&=
	\frac{{\mathrm e}^{-4 t}}{2}-{\mathrm e}^{-3 t}+\frac{1}{2}
		\end{split}
	\end{equation}
	
This lowest-order closure is quantitatively crude, but it preserves probability conservation and yields the lower bound $\pi_1(\infty)=1/2$. At this level, spatial correlations are neglected, and the asymptotic monomer and dimer densities are equal.
\subsection{Second order}
At second order, all probabilities associated with intervals of four or more sites are approximated by their mean-field forms. The closure requires the following three expressions:
\begin{equation}
	\begin{split}
		p_{0000}^{(2)}&=e^{-4t}\\
		p_{0010}^{(2)}&=e^{-3t}(1-e^{-t})\\
		p_{00010}^{(2)}&=e^{-4t}(1-e^{-t})
	\end{split}
\end{equation}

Solving the equation of $ p_{000}^{(2)}$ of Eq.~\eqref{eq:empty3},  one obtains
\begin{equation}
		p_{000}^{(2)}=-\frac{\left(3 {\mathrm e}^{-4 t}-8 {\mathrm e}^{-3 t}-1\right) {\mathrm e}^{-t}}{6}
\end{equation}
where the superscript $(2)$ denotes the second-order approximation. The long-time decay of $p_{000}^{(2)}$ is proportional to $e^{-t}$, in contrast with the faster $e^{-3t}$ decay obtained at first order. The approximations $p_{0010}^{(1)}$ and $p_{0010}^{(2)}$ are identical.

Integrating Eqs.~\eqref{eq:riviera} and \eqref{eq:md}, one obtains
\begin{equation}
	\begin{split}
		\pi_0(t)&= -\frac{{\mathrm e}^{-5 t}}{10}-\frac{{\mathrm e}^{-4 t}}{6}+\frac{2 {\mathrm e}^{-3 t}}{3}+\frac{{\mathrm e}^{-t}}{6}+\frac{13}{30}
		\\
		\pi_m(t)&=\frac{{\mathrm e}^{-5 t}}{10}-\frac{5 {\mathrm e}^{-4 t}}{6}+\frac{2 {\mathrm e}^{-3 t}}{3}-\frac{{\mathrm e}^{-t}}{6}+\frac{7}{30}
		\\
		\pi_d(t)&= \frac{{\mathrm e}^{-4 t}}{2}-\frac{2 {\mathrm e}^{-3 t}}{3}+\frac{1}{6}\\
		\pi_1(t)&=\frac{{\mathrm e}^{-5 t}}{10}+\frac{{\mathrm e}^{-4 t}}{6}-\frac{2 {\mathrm e}^{-3 t}}{3}-\frac{{\mathrm e}^{-t}}{6}+\frac{17}{30}
\end{split}\end{equation}

The second-order closure modifies the three-site probability and produces a slower long-time relaxation of $\pi_0(t)$ and $\pi_1(t)$ than at first order. It also lifts the degeneracy between the monomer and dimer densities. Although the resulting estimate of the jamming density is improved, it remains far from the numerical value quoted above.

\subsection{Third order}
At third order, only probabilities associated with intervals of five and six sites are replaced by mean-field expressions. Five such probabilities are required:
\begin{equation}
	\begin{split}
		p_{00000}^{(3)}&=e^{-5t}\\
		p_{000010}^{(3)}&=e^{-5t}(1-e^{-t})\\
		p_{00010}^{(3)}&=e^{-4t}(1-e^{-t})\\
		p_{000100}^{(3)}&=e^{-5t}(1-e^{-t})\\
		p_{010010}^{(3)}&=e^{-4t}(1-e^{-t})^2
	\end{split}
\end{equation}

Substitution into Eq.~\eqref{eq:empty3} gives 
\begin{equation}
	\begin{split}
p_{000}^{(3)}&=\frac{\left(-6 {\mathrm e}^{-5 t}+5 {\mathrm e}^{-4 t}+20 {\mathrm e}^{-3 t}+10 {\mathrm e}^{-t}+1\right) {\mathrm e}^{-t}}{30}\\
p_{0010}^{(3)}&=\frac{\left(9 {\mathrm e}^{-5 t}-40 {\mathrm e}^{-4 t}+30 {\mathrm e}^{-3 t}-5 {\mathrm e}^{-t}+6\right) {\mathrm e}^{-t}}{30}
	\end{split}
\end{equation}
Finally, the occupation probabilities are given by

\begin{equation}
	\begin{split}
\pi_0(t)&=\frac{{\mathrm e}^{-6 t}}{15}-\frac{{\mathrm e}^{-5 t}}{2}+\frac{2 {\mathrm e}^{-4 t}}{3}+\frac{13 {\mathrm e}^{-t}}{30}+\frac{1}{3}\\
\pi_1(t)&=-\frac{{\mathrm e}^{-6 t}}{15}+\frac{{\mathrm e}^{-5 t}}{2}-\frac{2 {\mathrm e}^{-4 t}}{3}-\frac{13 {\mathrm e}^{-t}}{30}+\frac{2}{3}
	\end{split}
\end{equation}

The third-order approximation produces a substantial change in the saturation density. More generally, the absolute error relative to the numerical estimate decreases as the truncation order increases, although the convergence exhibits oscillations with period three. This structure is associated with the successive inclusion of interval probabilities containing an additional occupied site.

\begin{figure}[H]
	\begin{center}
		\includegraphics[scale=0.45]{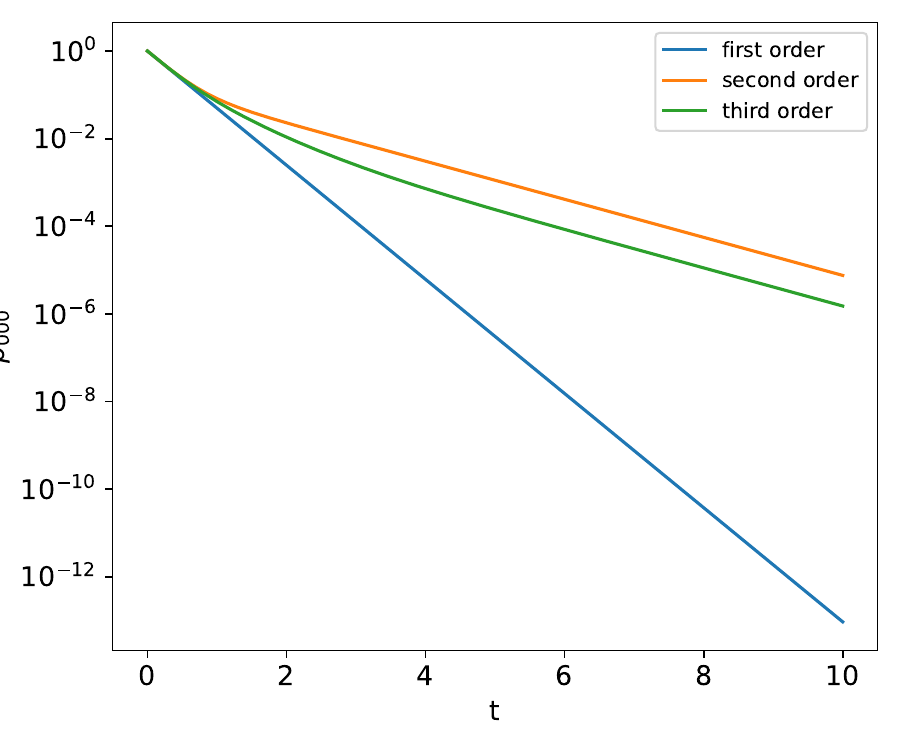}
		\includegraphics[scale=0.45]{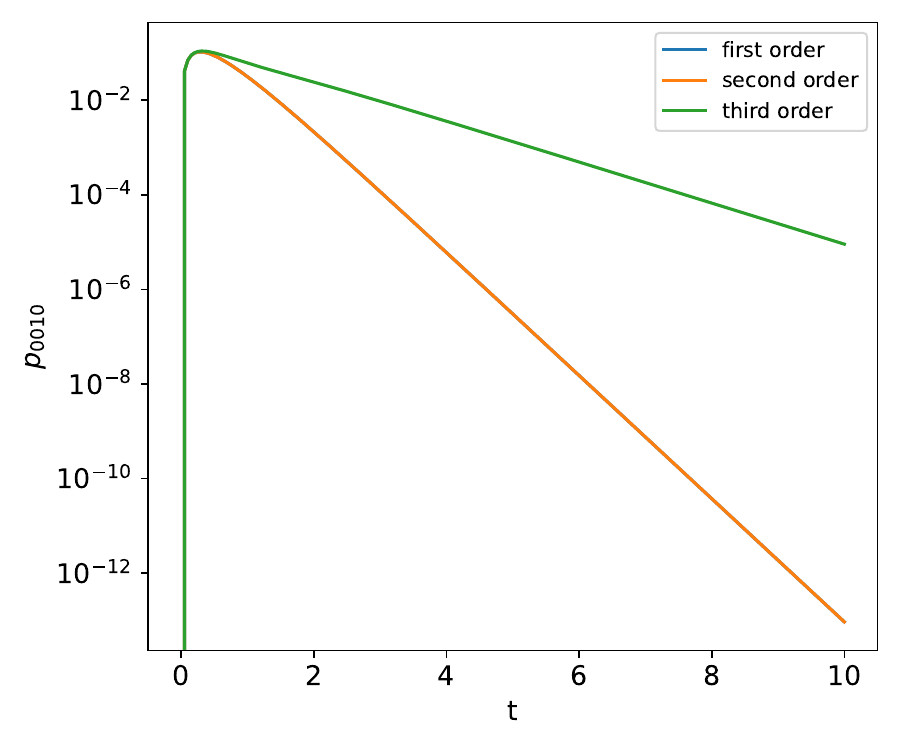}
	\end{center}
	\caption{Semilogarithmic plot of the time evolution of  $p_{000}(t)$ (left) and $p_{0010}(t)$ (right)  for the first three approximation orders.} \label{fig:p000p0010lo}
\end{figure}

Figure~\ref{fig:p000p0010lo} compares the first three approximation orders for the key quantities $p_{000}(t)$ and $p_{0010}(t)$. Relative to the first-order results, the third-order probabilities display significant long-time corrections, and both decay as $e^{-t}$. (Note that the two first orders of $p_{0010}(t)$ are identical). Their convergence is expected to improve rapidly as probabilities associated with longer intervals are incorporated.

 \section{Construction of the kinetic equations}
To generate the hierarchy systematically at arbitrary order, one must identify all joint probabilities that enter the kinetics and formulate an algorithm for constructing their evolution equations.

Consider an interval of length $l$. The relevant sequences have empty boundary sites. Within the interior of the interval, occupied sites must be separated in a manner compatible with the Riviera deposition rule. Left--right symmetry is used throughout to eliminate redundant sequences.

\begin{table}[H] 
	\centering
	
	\begin{tabular}{|l|l|l|}
		\hline
		length & number& \\
		\hline
		3&1& $p_{000}$\\	\hline
		4 &2 &$p_{0000}$, $p_{0010}$\\	\hline
		5& 3& $p_{00000}$, $p_{00010}$, $p_{00100}$\\	\hline
		6&4&$p_{000000}$, $p_{000010}$, $p_{000100}$,$p_{010010}$\\	\hline
		7&6&  $p_{0000000}$, $p_{0000010}$,$p_{0000100}$, $p_{0001000}$,
		$p_{0100010}$,$p_{0010010}$\\	\hline
	\end{tabular}
	\caption{Joint occupation probabilities required for intervals of lengths $l=3,\ldots,7$.}
	\label{table:1}
\end{table}

Table~\ref{table:1} lists the intervals that occur in the kinetic equations for several values of $l$.

 For $l=3$, only the fully empty sequence occurs. For $l=4$, the admissible intervals are the fully empty sequence and a sequence containing one occupied site.
 Intervals containing two occupied sites first appear at length $6$. Similarly, three occupied sites first occur at length $9$, and four occupied sites at length $12$. 
 
 The number of required intervals grows rapidly with the truncation order. For example, $k=24$, corresponding to the 22nd-order approximation, involves $2978$ distinct intervals.

 The evolution equation for a given interval of length $l$ is assembled from the following contributions. First, bulk depositions produce a loss term proportional to the interval probability multiplied by the number of admissible empty target sites in its interior. Second, depositions involving either boundary generate loss terms associated with intervals extended by an empty site. Third, when the second or penultimate site is empty, additional boundary losses involve intervals extended by the patterns $10$ or $01$. Finally, gain terms arise from longer configurations in which one occupied site is replaced by an empty site before deposition.

 To carry out the calculation at high order, we developed a Python program based on the SymPy computer-algebra library \cite{10.7717/peerj-cs.103}. The program enumerates the required intervals, constructs the corresponding differential equations, and obtains their exact solutions at arbitrary truncation order.

 \section{High-order approximations}
The symbolic implementation yields exact expressions for all joint probabilities once the mean-field closure is specified for intervals of length $k$ and greater.

\begin{figure}
	\begin{center}
		\includegraphics[scale=0.45]{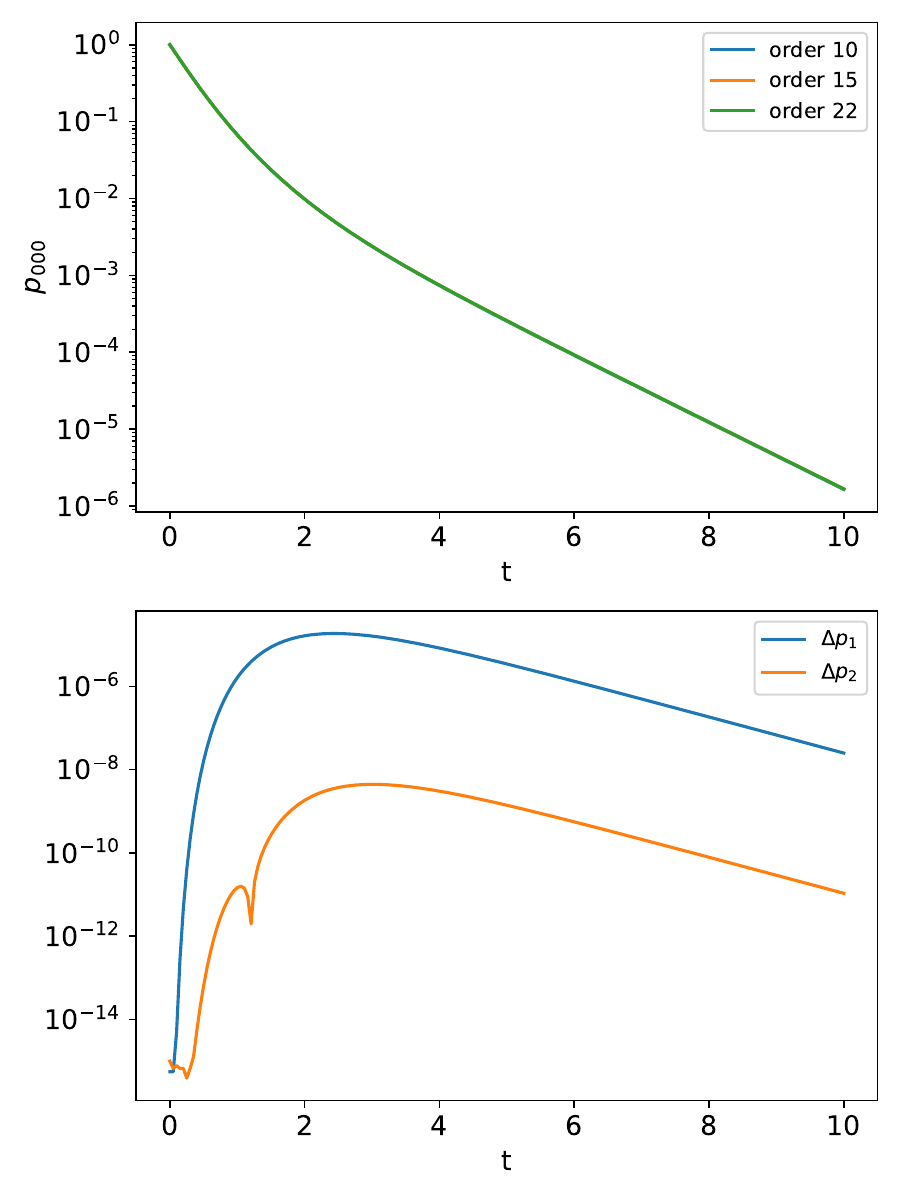}
		\includegraphics[scale=0.45]{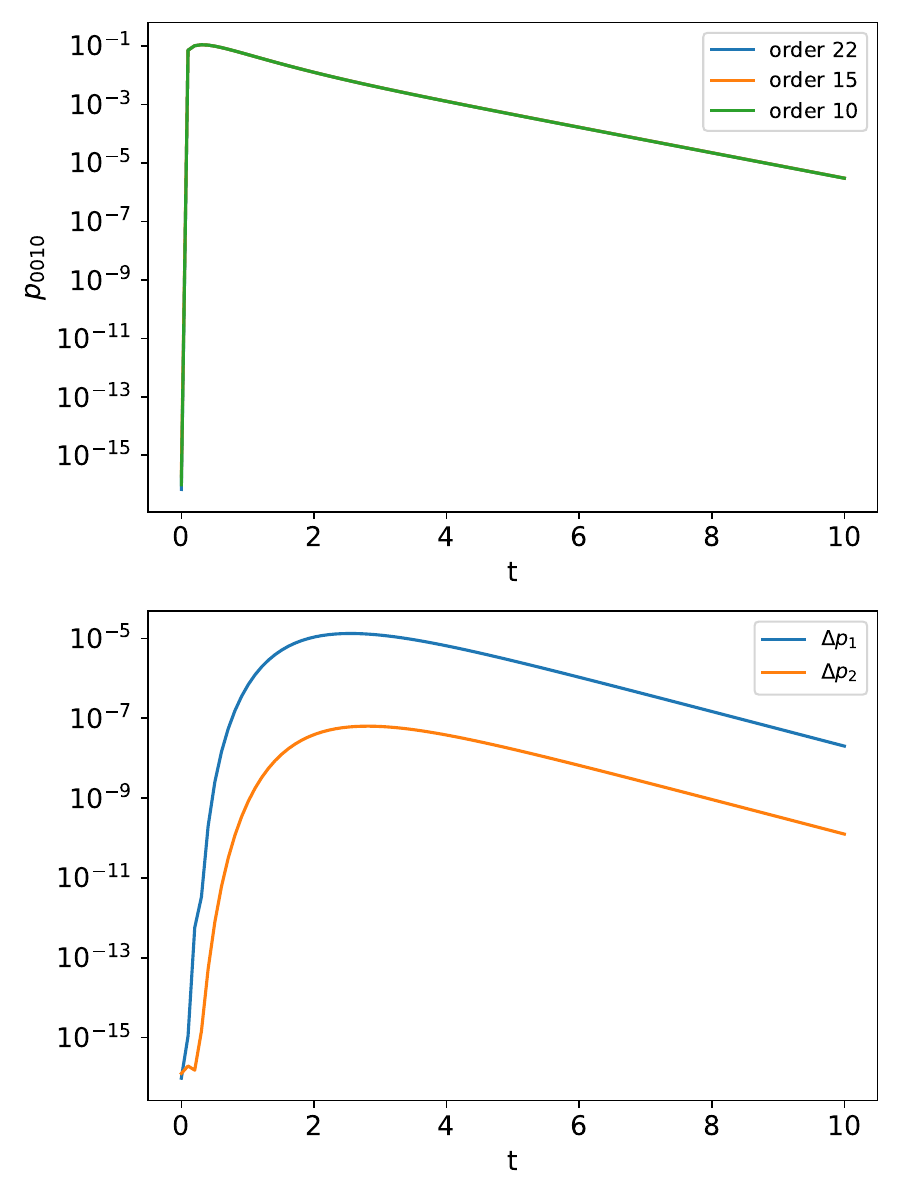}
	\end{center}
	\caption{Semilogarithmic plot of the time evolution of  $p_{000}(t)$ (top left) and $p_{0010}(t)$ (top right)  for the $10,15$ and $22$nd-order approximations. The absolute difference $\Delta p_1=|p^{(10)}-p^{(15)}|$ and $\Delta p_2=|p^{(15)}-p^{(22)}|$ shows the convergence
	 of the method for $p_{000}$ (bottom left) and $p_{0010}$ (bottom right). } \label{fig:p000p0010}
\end{figure}

 Figure~\ref{fig:p000p0010} compares the time evolution of $p_{000}(t)$ and $p_{0010}(t)$ at three  truncation orders.
 The corresponding curves (top panel) are visually indistinguishable, as are those obtained at still higher orders. Their maximum difference occurs at intermediate times and and decreases when the order of approximation increases.
 By computing the absolute difference between the $10$th order and the $15$th order , the maximum is  below $10^{-5}$, and between $15$th order and the $22$th order, the maximum is below $10^{-7}$. Because all coefficients of the expressions of the probabilities are obtained exactly as rational numbers, the convergence of the hierarchy can be quantified without numerical integration error.

Probabilities associated with longer intervals are not displayed because their number increases rapidly and they show the same convergence pattern.
 
 	\begin{figure}
 	\begin{center}
 		\includegraphics[scale=0.8]{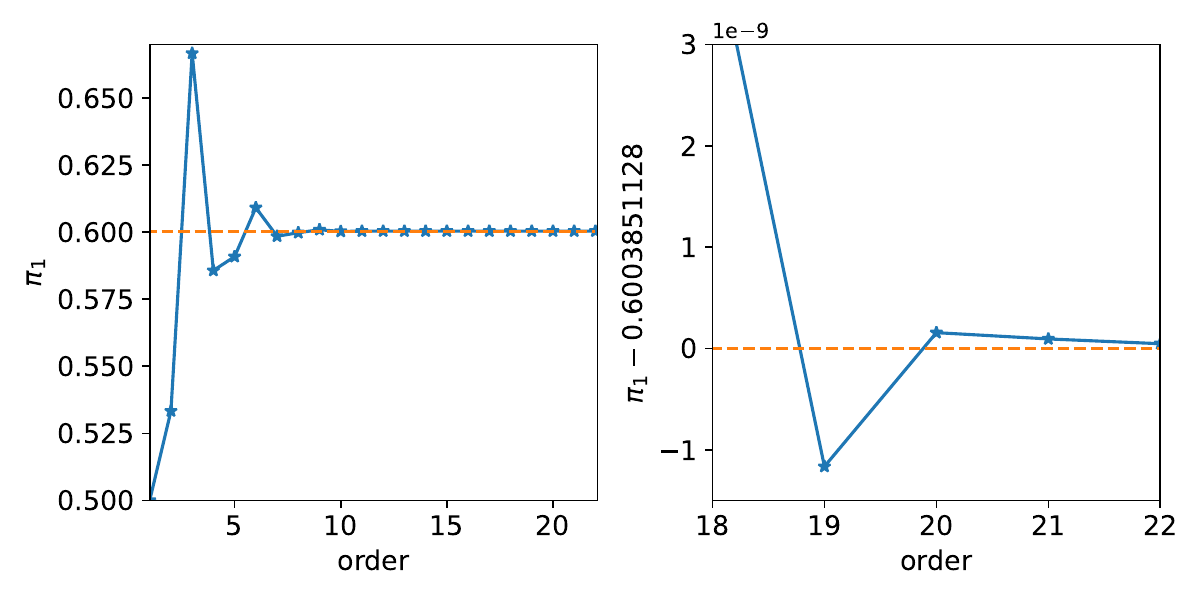}
 	\end{center}
 	\caption{Convergence of the occupied-site density as a function of the approximation order.}\label{fig:conver}
 \end{figure}	

Figure~\ref{fig:conver}(a) shows the jamming density as a function of the truncation order. Beyond the first few approximations, the sequence converges rapidly as correlations between deposited houses are incorporated over progressively larger distances. The convergence exhibits oscillations with period three. Each local extremum is associated with the first appearance of intervals containing an additional occupied site: two occupied sites become possible at length $6$, three at length $9$, and so forth. For $k>12$, the estimates are already extremely close to the numerical value reported in \cite{Krapivsky2023a}.

To show the high-order behavior, Fig.~\ref{fig:conver}(b) displays the difference between the extrapolated limiting value and the estimates obtained at successive truncation orders.  
Using the 22nd-order approximation, the long-time limit obtained for $ \pi_1$ is
\begin{equation}
	\begin{split}
	\pi_1&=0.6003851128\\
	\pi_m&=0.1252546914\\
	\pi_d&=0.2375652107\\
	\end{split}
\end{equation}

These values are consistent with the simulation results of \cite{Krapivsky2023a}, while providing several additional significant digits. The complete 22nd-order calculation requires approximately 30 minutes on a standard laptop.

 \begin{table}
 	\begin{center}
 		\begin{tabular}{|c|c|c|c||l|}
 			\hline
 			order  &$\pi_m$&$\pi_d$&$\pi_1$ & \\
 			\hline\hline
 			1&$\frac{1}{6}$&$\frac{1}{6}$& $\frac{1}{2}$&0.5    \\
 			\hline\hline
 			2&$\frac{7}{30}$&$\frac{1}{6}$& $\frac{13}{30} $&0.53    \\
 			\hline
 			3 &$\frac{1}{15}$&$\frac{3}{10}$&$\frac{2}{3}$&0.666 \\
 			\hline
 			4 &$\frac{149}{630}$&$\frac{71}{630}$&$\frac{41}{70}$&0.58571 \\
 			\hline
 			5 &$\frac{359}{2520}$&$\frac{113}{504}$& $\frac{1489}{2520}$& 0.590873 \\
 			\hline
 			6&$\frac{16}{135}$ &$\frac{3709}{15120}$& $\frac{307}{504}$& 0.6091267\\
 			\hline
 			7&$\frac{3539}{28350}$&$\frac{107423}{453600}$&$\frac{9049}{15120}$  & 0.5984789 \\
 			\hline
 			8 &$\frac{157739}{1247400}$&$\frac{1180877}{4989600}$& $\frac{14251}{23760}$ &0.5997895623\\
 			\hline
 			9&$ \frac{28307}{226800}$&$\frac{2375501}{9979200}$& $\frac{599651}{997920}$&0.600900974  \\
 			\hline
 			10&$ \frac{12188923}{97297200}$&$\frac{184861909}{778377600} $& $\frac{33373943}{55598400}$&0.60026804728193\\
 			\hline
 			11&$\frac{34134301}{272432160}$&$\frac{2588518601}{10897286400}$& $\frac{3271204621}{5448643200}$&0.6003704960897421\\
 			\hline
 			12 &$\frac{30463463}{243243000}$&$\frac{12944963371}{54486432000}$&$\frac{5452290409}{9081072000}$&0.6004016275831752
 			\\
 			\hline
 			13 &$\frac{787473733}{6286896000} $&$\frac{144088229}{606528000} $ &$\frac{32712624163}{54486432000}$&0.6003811033726708
 			
 			\\
 			\hline
 			14 &$\frac{19889188289}{158789030400} $&$\frac{10562351459993}{44460928512000} $&$\frac{4448945940151}{7410154752000}$&0.6003850242061719
 			\\
 			\hline
 			15 &$\frac{4176693405329}{33345696384000}$&$\frac{63374308056041}{266765571072000}$&$\frac{80081081677357}{133382785536000}$& 0.6003854347137106 	  
 			\\
 			\hline
 			16&$\frac{29759034287597}{237588086736000}$&$\frac{190122603244183}{800296713216000}$&$\frac{4564618558842581}{7602818775552000}$ &0.60038502739547 
 			\\
 			\hline
 			17&$
 			\frac{2380721943968353}{19007046938880000}$&$\frac{36123304562861221}{152056375511040000}$&$\frac{45646192338734633}{76028187755520000}$&0.6003851161823925
 			\\
 			\hline
 			18&$ \frac{33330103872966539}{266098657144320000}$&$ \frac{505726278059248567}{2128789257154560000}$&$\frac{30430794931005463}{50685458503680000}$&0.6003851169422892\\
 			\hline
 			19&$\frac{199980632471710511}{1596591942865920000}$&$ \frac{33377933572762255709}{140500090972200960000}$&$\frac{14059027133839172731}{2341668182870016000}$	&0.6003851116347332\\
 			\hline
 			20&$\frac{25297549840176809771}{201968880772538880000}$&$\frac{767692475647377498137}{3231502092360622080000}$&$\frac{21557174985973155029}{35905578804006912000}$&0.6003851129554125\\
 			\hline
 			21 &$\frac{33730066431823831519}{269291841030051840000} $&$\frac{102358996757030127743}{430866945648082944000}$&$\frac{646715249512445964791}{1077167364120207360000}$&0.600385112893446
 			\\
 			\hline
 			22&$\frac{250884791758990075783}{2002997164686336000000} $&$\frac{20026760230610602962503}{84300054583320576000000} $&$\frac{582043724515395064816541}{969450627708186624000000} $&0.6003851128461959\\
 			\hline
 			simu&\cite{Krapivsky2023a}&&&0.600385\\ 
 			\hline
 		\end{tabular}
 	\end{center}
 	\caption{Asymptotic monomer density $\pi_m$, dimer density $\pi_d$, and occupied-site density $\pi_1$ at successive approximation orders. The last column gives the decimal value of $\pi_1$.}
	\label{table:densities}
 \end{table}

Table~\ref{table:densities} summarizes the monomer, dimer, and total occupied-site densities obtained through 22 approximation orders. Owing to the linear structure of the truncated hierarchy and to the exponential mean-field closure, all coefficients remain rational. Alternative closures are possible, but they generally generate special functions for probabilities associated with long intervals and thereby destroy the simple exact solvability of the finite systems.

A physical interpretation of the rapid convergence can be related to the decay of the correlation function, which was studied by Krapivsky and Luck \cite{Krapivsky2023b}. They showed that the correlations decay exponentially, accompanied by oscillations with period 3, with a characteristic length of approximately $1.23$. At higher orders, where the closure involves joint probabilities over lengths $l$ much larger than the correlation length, the effect of the approximation progressively vanishes.

\section{Kinetics of the model}
Once the joint probabilities are known, a variety of time-dependent observables can be evaluated. At the highest truncation orders, the resulting dynamics is sufficiently accurate to provide a detailed description of the approach to jamming. 
	\begin{figure}
	\begin{center}
		\includegraphics[scale=0.7]{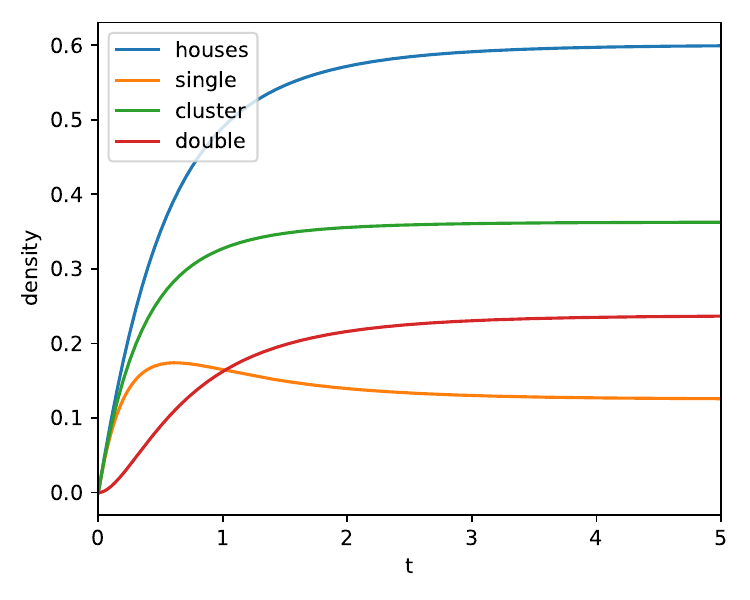}
	\end{center}
	\caption{Time evolution of the monomer density $\pi_m(t)$, dimer density $\pi_d(t)$, cluster density $\pi_{cl}(t)$, and total occupied-site density $\pi_1(t)$.}
\label{fig:kinetics}
\end{figure}

Using the highest-order approximation, we obtain highly accurate expressions for the time-dependent monomer, dimer, cluster, and occupied-site densities. Explicit formulas for $\pi_m(t)$, $\pi_d(t)$, and $\pi_1(t)$ are given in ~\ref{sec:mdappendix}.

Figure~\ref{fig:kinetics} shows their evolution from an initially empty lattice. The total occupied-site density and the dimer density increase monotonically, whereas the monomer density exhibits a pronounced maximum at finite time $t=0.63$. This nonmonotonic behavior has a simple kinetic interpretation. At short times, most successful depositions occur in large empty regions and therefore create isolated houses. At intermediate times, new houses are increasingly deposited next to existing monomers, converting them into dimers. This behavior is illustrated in Fig.~\ref{fig:kinetics}, where the cluster density reaches a plateau well before the other densities attain their saturation values.  The monomer density consequently decreases while the dimer density continues to grow.
The monomer and dimer densities cross at a time close to $t=1$, marking the onset of the kinetic regime in which the remaining available regions are essentially isolated target sites.

\section{Conclusion}

We have developed a systematic analytical framework for the kinetics of the one-dimensional Riviera model, a remarkably simple irreversible deposition process that nevertheless evades an exact finite-dimensional solution because its dynamics generates correlations over arbitrarily long distances. Starting from the exact hierarchy of joint occupation probabilities, we introduced a sequence of finite closures in which probabilities associated with sufficiently long intervals are replaced by mean-field expressions. Each truncation produces a linear system that can be solved exactly and yields explicit formulas for the full temporal evolution.

The most striking result is the rapid convergence of this hierarchy with increasing truncation order. The lowest-order closures provide only a qualitative description, but the inclusion of progressively longer intervals rapidly restores the spatial correlations generated by the deposition dynamics. At high order, the estimates of the jamming density stabilize to many significant digits. 

The method provides considerably more information than the final jamming density alone. Because the complete time dependence remains analytically accessible, one can follow separately the formation and subsequent conversion of monomers and dimers during the approach to saturation. In particular, the nonmonotonic evolution of the monomer density reveals the crossover from deposition in extended empty regions to deposition adjacent to previously built houses. The hierarchy therefore offers a detailed kinetic picture of how the jammed state is assembled.

More broadly, this work demonstrates that the absence of an exact finite closure need not preclude an essentially analytical treatment of a correlated irreversible process. Solvable kinetic hierarchies combined with systematically improvable truncations provide an efficient alternative to direct simulation when standard interval methods fail. The Riviera model is an especially transparent example, but the same strategy should be applicable to other one-dimensional constrained adsorption models, to modified settlement rules involving extended objects or heterogeneous deposition rates, and potentially to more complex geometries. Establishing quantitative convergence criteria for such hierarchy closures would be a valuable direction for future work.

\section*{Acknowledgements}
I warmly thank P.~L. Krapivsky and J.-M. Luck for many stimulating discussions and for introducing me to the Riviera model. I also thank L. Gerin for valuable discussions on the mathematical aspects of the model.

\appendix
\section{Occupation probabilities at the highest order}
\label{sec:mdappendix}
The occupation probabilities $\pi_m(t)$, $\pi_d(t)$, and $\pi_1(t)$ are given below at 22nd order. At approximation order $n$, the number of exponential terms grows as $n+1$ for $\pi_m(t)$ and $\pi_d(t)$, and as $n+3$ for $\pi_1(t)$.
\begin{equation}
	\begin{split}
		\pi_d(t)&=\frac{767692475647377498137}{3231502092360622080000} - \frac{563534595228307657 e^{- t}}{4257578514309120000} - \frac{788559397182242393 e^{- 2 t}}{4257578514309120000}\\& - \frac{139739674599323 e^{- 3 t}}{2824576634880000} + \frac{49540011925783 e^{- 4 t}}{1501791363072000} + \frac{1858601849973851 e^{- 5 t}}{32011868528640000}\\
		& + \frac{1051218343982441 e^{- 6 t}}{32011868528640000} + \frac{5073885326971 e^{- 7 t}}{549223234560000} - \frac{1926102215669 e^{- 8 t}}{2196892938240000} \\
		&- \frac{7589825037029 e^{- 9 t}}{3954407288832000} - \frac{701290189681 e^{- 10 t}}{941525544960000} - \frac{78312150821 e^{- 11 t}}{398337730560000}\\
		& + \frac{20456290711 e^{- 12 t}}{217275125760000} - \frac{8632330867 e^{- 13 t}}{313841848320000} + \frac{53957574763 e^{- 14 t}}{1318135762944000}\\
		& - \frac{29627441573 e^{- 15 t}}{1647669703680000} + \frac{31858967587 e^{- 16 t}}{4393785876480000} - \frac{25288052107 e^{- 17 t}}{10670622842880000}\\
		& + \frac{31117154213 e^{- 18 t}}{96035605585920000} + \frac{3219148721 e^{- 19 t}}{60822550204416000} - \frac{1684801273 e^{- 20 t}}{35777970708480000}\\
		& + \frac{187396763929 e^{- 21 t}}{12772735542927360000} - \frac{1768239349 e^{- 22 t}}{821637958901760000} + \frac{718637809 e^{- 23 t}}{6120269114319360000}
			\end{split}
	\end{equation}
	\begin{equation}
		\begin{split}
		\pi_m(t)=&\frac{25297549840176809771}{201968880772538880000} + \frac{839183918977563109 e^{- t}}{8781255685762560000} + \frac{324198727588831 e^{- 2 t}}{3942202328064000}\\
		& - \frac{2306936435907497 e^{- 3 t}}{28510570408320000} - \frac{1260725076660431 e^{- 4 t}}{11404228163328000} - \frac{55012942198619 e^{- 5 t}}{666913927680000}\\
		& - \frac{125244770144423 e^{- 6 t}}{4001483566080000} - \frac{407878769 e^{- 7 t}}{99018612000} + \frac{152553652297 e^{- 8 t}}{45768602880000}\\
		& + \frac{576910352797 e^{- 9 t}}{247150455552000} + \frac{239640650167 e^{- 10 t}}{353072079360000} + \frac{98509633 e^{- 11 t}}{1383117120000}\\
		& - \frac{339518857 e^{- 12 t}}{3734416224000} + \frac{964084117 e^{- 13 t}}{176536039680000} - \frac{1544975171 e^{- 14 t}}{54922323456000} \\
		&+ \frac{177586547 e^{- 15 t}}{12872419560000} - \frac{4721641129 e^{- 16 t}}{823834851840000} + \frac{97336297 e^{- 17 t}}{44460928512000}\\
		& - \frac{4920015389 e^{- 18 t}}{12004450698240000} - \frac{4509977 e^{- 19 t}}{5702114081664000} + \frac{4936073 e^{- 20 t}}{162453392640000}\\
		& - \frac{9388636931 e^{- 21 t}}{798295971432960000} + \frac{1371523801 e^{- 22 t}}{702500454861004800} - \frac{82524181 e^{- 23 t}}{701280836015760000}		
	\end{split}
\end{equation}
\begin{equation}
	\begin{split}
	\pi_1(t)&=\frac{582043724515395064816541}{969450627708186624000000} - \frac{546625825268650585109 e^{- t}}{3231502092360622080000} - \frac{1277477466137187071 e^{- 2 t}}{4432787506667520000} \\&- \frac{75811252875526033193 e^{- 3 t}}{421500272916602880000} - \frac{25304383861454657 e^{- 4 t}}{567677135241216000} + \frac{102286232777793629 e^{- 5 t}}{3041127510220800000}\\& + \frac{26874809639316299 e^{- 6 t}}{782004216913920000} + \frac{743755429129271 e^{- 7 t}}{51711479930880000} + \frac{457068925672951 e^{- 8 t}}{298777439600640000} \\&- \frac{16884637380587 e^{- 9 t}}{11863221866496000} - \frac{91063945743013 e^{- 10 t}}{98860182220800000} - \frac{4754337914717 e^{- 11 t}}{24165822320640000} \\&- \frac{4046345629279 e^{- 12 t}}{186422057902080000} + \frac{66228661687 e^{- 13 t}}{1479540142080000} - \frac{113927498933 e^{- 14 t}}{14499493392384000} + \frac{3086123198629 e^{- 15 t}}{296580546662400000}\\& - \frac{742043073329 e^{- 16 t}}{158176291553280000} + \frac{109063022897 e^{- 17 t}}{74694359900160000} - \frac{815855671927 e^{- 18 t}}{2016747717304320000} \\&+ \frac{5498090033 e^{- 19 t}}{364935301226496000} + \frac{723310099 e^{- 20 t}}{22279322419200000} - \frac{566271078949 e^{- 21 t}}{38318206628782080000} \\&+ \frac{184073672789 e^{- 22 t}}{46833363657400320000} - \frac{622167522461 e^{- 23 t}}{1077167364120207360000} + \frac{251236951921 e^{- 24 t}}{7755605021665492992000}\\& + \frac{6960036683 e^{- 25 t}}{20196888077253888000000}	
	\end{split}
\end{equation}

\section*{References}

\begin{thebibliography}{10}
\expandafter\ifx\csname url\endcsname\relax
  \def\url#1{{\tt #1}}\fi
\expandafter\ifx\csname urlprefix\endcsname\relax\def\urlprefix{URL }\fi
\providecommand{\eprint}[2][]{\url{#2}}

\bibitem{Torquato10}
Torquato S 2010 Reformulation of the covering and quantizer problems as ground
  states of interacting particles {\em Phys. Rev. E\/}
  \href{http://dx.doi.org/10.1103/PhysRevE.82.056109}{{\bf 82}(5) 056109}

\bibitem{Chen2014}
Chen E~R, Klotsa D, Engel M, Damasceno P~F and Glotzer S~C 2014 Complexity in
  surfaces of densest packings for families of polyhedra {\em Phys. Rev. X\/}
  \href{http://dx.doi.org/10.1103/PhysRevX.4.011024}{{\bf 4}(1) 011024}

\bibitem{Jiao2009}
Jiao Y, Stillinger F~H and Torquato S 2009 Optimal packings of superballs {\em
  Phys. Rev. E\/} \href{http://dx.doi.org/10.1103/PhysRevE.79.041309}{{\bf
  79}(4) 041309}

\bibitem{Hales05}
Hales T~C 2005 A proof of the {K}epler conjecture {\em Ann. Math.\/} {\bf 168}
  1065--1185

\bibitem{doslic2020}
Došlić T, Taheri-Dehkordi M and Fath-Tabar G~H 2020 Packing stars in
  fullerenes {\em Journal of Mathematical Chemistry\/}
  \href{http://dx.doi.org/https://urn.nsk.hr/urn:nbn:hr:237:675944}{{\bf 58}
  2223--2244}

\bibitem{Conway}
Conway J~H and Sloane N~J~A 1999 {\em Sphere Packings, Lattices and Groups\/}
  (New York: Springer-Verlag)
  \urlprefix\url{https://doi.org/10.1007/978-1-4757-6568-7}

\bibitem{Ewens2005}
Ewens W~J and Grant G 2005 {\em Statistical Methods in Bioinformatics An
  Introduction\/} (2nd ed. New York, Springer Science)

\bibitem{Athreya04}
Athreya S, Roy R and Sarkar A 2004 On the coverage of space by random sets {\em
  Advances in Applied Probability\/}
  \href{http://dx.doi.org/http://www.jstor.org/stable/1428350}{{\bf 36} 1--18}
  ISSN 00018678

\bibitem{Bobrowski17}
Bobrowski O and Weinberger S 2017 On the vanishing of homology in random
  \v{C}ech complexes {\em Random Struct. Algorithms\/} {\bf 51} 14--51

\bibitem{Tillmann22}
de~Kergorlay H~L, Tillmann U and Vipond O 2022 Random \v{C}ech complexes on
  manifolds with boundary {\em Random Struct. Algorithms\/} {\bf 62} 1--44

\bibitem{Leykam2023}
Leykam D and Angelakis D~G 2023 Topological data analysis and machine learning
  {\em Advances in Physics: X\/}
  \href{http://dx.doi.org/10.1080/23746149.2023.2202331}{{\bf 8} 2202331}
  (\textit{Preprint} \eprint{https://doi.org/10.1080/23746149.2023.2202331})

\bibitem{Hall88}
Hall P 1988 {\em Introduction to the Theory of Coverage Processes\/} (New York:
  Wiley)

\bibitem{Rogers}
Rogers C~A 1964 {\em Packing and Covering\/} (Cambridge, UK: Cambridge
  University Press)

\bibitem{GL87}
Gruber P~M and Lekkerkerker C~G 1987 {\em Geometry of numbers\/} (Amsterdam:
  North-Holland)

\bibitem{Kannan1987}
Kannan R 1987 Algorithmic geometry of numbers {\em Annual Review of Computer
  Science\/}
  \href{http://dx.doi.org/https://doi.org/10.1146/annurev.cs.02.060187.001311}{{\bf
  2} 231--267} ISSN 8756-7016

\bibitem{ZZ2008}
Zhigljavsky A and \u{Z}ilinskas A 2008 {\em Stochastic Global Optimization\/}
  (New York: Springer) ISBN 978-0-387-74022-5
  \urlprefix\url{https://doi.org/10.1007/978-0-387-74740-8}

\bibitem{Gomez2019}
Gomez J 2019 Stochastic global optimization algorithms: A systematic formal
  approach {\em Information Sciences\/}
  \href{http://dx.doi.org/https://doi.org/10.1016/j.ins.2018.09.021}{{\bf 472}
  53--76}

\bibitem{Puljiz2022}
Puljiz M, Šebek S and Žubrinić J 2022 Combinatorial settlement planning {\em
  Contribution to discrete Mathematics\/}
  \href{http://dx.doi.org/https://doi.org/10.55016/ojs/cdm.v18i2.73491}{{\bf
  18} 20} ISSN 1715-0868

\bibitem{Puljiz2023}
Puljiz M, Šebek S and Žubrinić J 2023 Packing density of combinatorial
  settlement planning models {\em The American Mathematical Monthly\/}
  \href{http://dx.doi.org/10.1080/00029890.2023.2254181}{{\bf 130} 915--928}
  ISSN 0002-9890

\bibitem{Puljiz2025}
DO\v{S}LIĆ T, PULJIZ M, \v{S}EBEK S and \v{Z}UBRINIĆ J 2025 Predators and
  altruists arriving on jammed riviera {\em Advances in Complex Systems\/}
  \href{http://dx.doi.org/10.1142/S0219525925500109}{{\bf 28} 2550010}
  (\textit{Preprint} \eprint{https://doi.org/10.1142/S0219525925500109})

\bibitem{Spahn2025}
Spahn G and Zeilberger D 2025 Enumerating seating arrangements that obey social
  distancing {\em Journal of Symbolic Computation\/}
  \href{http://dx.doi.org/https://doi.org/10.1016/j.jsc.2024.102344}{{\bf 126}
  102344} ISSN 0747-7171

\bibitem{Doslic2024}
Došlić T, Puljiz M, Šebek S and Žubrinić J 2024 On a variant of flory
  model {\em Discrete Applied Mathematics\/}
  \href{http://dx.doi.org/https://doi.org/10.1016/j.dam.2024.06.011}{{\bf 356}
  269--292} ISSN 0166-218X

\bibitem{Luck2024}
Luck J~M 2024 On the structure factor of jammed particle configurations on the
  one-dimensional lattice {\em J. Phys. A: Math. Theor.\/}
  \href{http://dx.doi.org/10.1088/1751-8121/ad469c}{{\bf 57} 225002}

\bibitem{Flory39}
Flory P~J 1939 Intramolecular reaction between neighboring substituents of
  vinyl polymers {\em J. Amer. Chem. Soc.\/}
  \href{http://dx.doi.org/10.1021/ja01875a053}{{\bf 61} 1518--1521}

\bibitem{Renyi58}
R{\'e}nyi A 1958 On a one-dimensional random space-filling problem {\em Publ.
  Math. Inst. Hung. Acad. Sci.\/} {\bf 3} 109--127

\bibitem{Evans93}
Evans J~W 1993 Random and cooperative sequential adsorption {\em Rev. Mod.
  Phys.\/} \href{http://dx.doi.org/10.1103/RevModPhys.65.1281}{{\bf 65}(4)
  1281--1329}

\bibitem{Talbot2000}
Talbot J, Tarjus G, {Van Tassel} P and Viot P 2000 From car parking to protein
  adsorption: an overview of sequential adsorption processes {\em Colloids
  Surfaces A\/}
  \href{http://dx.doi.org/https://doi.org/10.1016/S0927-7757(99)00409-4}{{\bf
  165} 287--324} ISSN 0927-7757

\bibitem{Krapivsky2026}
Krapivsky P~L 2026 Riviera model with egoistical settlers  (\textit{Preprint}
  \eprint{2606.16791})

\bibitem{Cohen1963}
Cohen E~R and Reiss H 1963 Kinetics of reactant isolation. i. one‐dimensional
  problems {\em The Journal of Chemical Physics\/}
  \href{http://dx.doi.org/10.1063/1.1733723}{{\bf 38} 680--691} ISSN 0021-9606
  (\textit{Preprint}
  \eprint{https://pubs.aip.org/aip/jcp/article-pdf/38/3/680/18828097/680_1_online.pdf})

\bibitem{Bonnier1994}
Bonnier B, Boyer D and Viot P 1994 Pair correlation function in random
  sequential adsorption processes {\em J. Phys. A: Math. Gen.\/} {\bf 27} 3671

\bibitem{Redner2010}
Krapivsky P~L, Redner S and Ben-Naim E 2010 {\em A Kinetic View of Statistical
  Physics\/} (Cambridge, UK: Cambridge University Press)
  \urlprefix\url{https://doi.org/10.1017/CBO9780511780516}

\bibitem{Krapivsky2023a}
Krapivsky P~L and Luck J~M 2023 A renewal approach to configurational entropy
  in one dimension {\em J. Phys. A: Math. Theor.\/}
  \href{http://dx.doi.org/10.1088/1751-8121/acd5bd}{{\bf 56} 255001}

\bibitem{10.7717/peerj-cs.103}
Meurer A, Smith C~P, Paprocki M, \v{C}ert\'{i}k O, Kirpichev S~B, Rocklin M,
  Kumar A, Ivanov S, Moore J~K, Singh S, Rathnayake T, Vig S, Granger B~E,
  Muller R~P, Bonazzi F, Gupta H, Vats S, Johansson F, Pedregosa F, Curry M~J,
  Terrel A~R, Rou\v{c}ka v, Saboo A, Fernando I, Kulal S, Cimrman R and Scopatz
  A 2017 Sympy: symbolic computing in python {\em PeerJ Computer Science\/}
  \href{http://dx.doi.org/10.7717/peerj-cs.103}{{\bf 3} e103} ISSN 2376-5992

\bibitem{Krapivsky2023b}
Krapivsky P~L and Luck J~M 2023 Jamming and metastability in one dimension:
  from the kinetically constrained ising chain to the riviera model {\em The
  European Physical Journal Special Topics\/} {\bf 232} 1703--1719 ISSN
  1951-6401

\end{thebibliography}

\providecommand{\newblock}{}

\end{document}